\documentclass[conference]{IEEEtran}
\IEEEoverridecommandlockouts
\usepackage[utf8]{inputenc} 
\usepackage[T1]{fontenc}
\usepackage{cite}
\usepackage{amsmath,amssymb,amsfonts}
\usepackage[T1]{fontenc}
\usepackage{xurl}             
\usepackage{nicefrac}
\usepackage{microtype}      
\usepackage{float}
\usepackage{booktabs}
\usepackage{listingsutf8}
\usepackage{CJKutf8}
\usepackage{algorithmic}
\usepackage{graphicx}
\usepackage{textcomp}
\usepackage{xcolor}
\usepackage{hyperref}
\usepackage{comment}
\usepackage{svg}
\usepackage{balance}
\usepackage{tikz}

\newcommand\copyrighttext{%
  \footnotesize \textcopyright 2025 IEEE. Personal use of this material is permitted. Permission from IEEE must be obtained for all other uses, including reprinting/republishing this material for advertising or promotional purposes, collecting new collected works for resale or redistribution to servers or lists, or reuse of any copyrighted component of this work in other works.}

\newcommand\copyrightnotice{%
\begin{tikzpicture}[remember picture,overlay]
\node[anchor=south,yshift=10pt] at (current page.south) {\fbox{\parbox{\dimexpr0.75\textwidth-\fboxsep-\fboxrule\relax}{\copyrighttext}}};
\end{tikzpicture}%
}

\hypersetup{colorlinks=true, urlcolor=blue, linkcolor=black}

\def\BibTeX{{\rm B\kern-.05em{\sc i\kern-.025em b}\kern-.08em
    T\kern-.1667em\lower.7ex\hbox{E}\kern-.125emX}}

\definecolor{vscodeBackground}{RGB}{255,255,255} % White background 
\definecolor{vscodeForeground}{RGB}{30,30,30} % Dark gray for general text 
\definecolor{vscodeKeyword}{RGB}{0,0,255} % Blue for keywords (e.g., def, import) 
\definecolor{vscodeComment}{RGB}{0,128,0} % Green for comments 
\definecolor{vscodeString}{RGB}{163,21,21} % Red for strings 
\definecolor{vscodeNumber}{RGB}{9,134,88} % Greenish for numbers 
\definecolor{vscodeOperator}{RGB}{0,0,0} % Black for operators 
\definecolor{vscodeIdentifier}{RGB}{0,0,0} % Black for identifiers/variable names 
\definecolor{vscodeFunction}{RGB}{72,118,214} % Blue for function names
\definecolor{backcolour}{RGB}{249,249,249}
\definecolor{framecolour}{RGB}{229, 229, 229}

\lstdefinestyle{mystyle}{
    backgroundcolor=\color{backcolour},   
    commentstyle=\color{vscodeComment},
    keywordstyle=\color{vscodeKeyword}\ttfamily,
    numberstyle=\tiny\color{vscodeNumber}\ttfamily,
    stringstyle=\color{vscodeString}\ttfamily,
    basicstyle=\footnotesize\ttfamily,
    breakatwhitespace=false,         
    breaklines=true,                 
    captionpos=b,                    
    keepspaces=true,                                  
    numbersep=3pt,                  
    showspaces=false,                
    showstringspaces=false,
    showtabs=false,
    columns=fullflexible,
    tabsize=2,
    frame=single,
    rulecolor=\color{framecolour}
}

\begin{document}
\bstctlcite{IEEEexample:BSTcontrol}

\title{Automated structural testing of LLM-based agents: \\methods, framework, and case studies}

\author{\IEEEauthorblockN{Jens Kohl\IEEEauthorrefmark{1},
Otto Kruse\IEEEauthorrefmark{2},
Youssef Mostafa\IEEEauthorrefmark{1}, 
Andre Luckow\IEEEauthorrefmark{1},
Karsten Schroer\IEEEauthorrefmark{2},
Thomas Riedl\IEEEauthorrefmark{1},\\
Ryan French\IEEEauthorrefmark{2},
David Katz\IEEEauthorrefmark{1},
Manuel P. Luitz\IEEEauthorrefmark{1},
Tanrajbir Takher\IEEEauthorrefmark{2},
Ken E. Friedl\IEEEauthorrefmark{1} and
Céline Laurent-Winter\IEEEauthorrefmark{1}}

\IEEEauthorblockA{\IEEEauthorrefmark{1}BMW Group, Munich, Germany}
\IEEEauthorblockA{\IEEEauthorrefmark{2}Amazon Web Services}}

\maketitle

\copyrightnotice

\begin{abstract}
LLM-based agents are rapidly being adopted across diverse domains. Since they interact with users without supervision, they must be tested extensively. Current testing approaches focus on acceptance-level evaluation from the user's perspective. While intuitive, these tests require manual evaluation, are difficult to automate, do not facilitate root cause analysis, and incur expensive test environments. In this paper, we present methods to enable structural testing of LLM-based agents. Our approach utilizes traces (based on OpenTelemetry) to capture agent trajectories, employs mocking to enforce reproducible LLM behavior, and adds assertions to automate test verification. This enables testing agent components and interactions at a deeper technical level within automated workflows. We demonstrate how structural testing enables the adaptation of software engineering best practices to agents, including the test automation pyramid, regression testing, test-driven development, and multi-language testing. In representative case studies, we demonstrate automated execution and faster root-cause analysis. Collectively, these methods reduce testing costs and improve agent quality through higher coverage, reusability, and earlier defect detection. We provide an open source reference implementation on \href{https://github.com/awslabs/generative-ai-toolkit}{GitHub}.
\end{abstract} 

\begin{IEEEkeywords}
Large language models, Intelligent agents,
Software agents, Automatic testing, Glass box.% https://www.ieee.org/publications/services/thesaurus-thank-you
\end{IEEEkeywords}

\section{Introduction} \label{sec:intro}

Agentic AI represents a paradigm in which autonomous LLM-based agents dynamically perceive, plan, and execute complex tasks by selecting appropriate tools, databases, and API. Unlike traditional hard-coded workflows, agents autonomously determine action sequences based on task context. For instance, an agent handling a customer request might query a database, analyze retrieved data using an LLM, and generate a tailored response without predefined instructions. These systems promise to transform productivity across domains (\cite{yee2024agents},\cite{gartner2024top}), with growing adoption in customer support and software engineering \cite{jin2024llms}.

%From perspective of a conceptual framework, agents can be structured in three modules. Serving as controller, the brain module undertakes basic tasks like memorizing, thinking, and decision-making. The perception module perceives and processes multimodal information from the external environment, and the action module carries out the execution using tools and influences the surroundings (\cite[Chap. 3]{xi2023rise}, more details in Sec. \ref{sec:relatedwork}).

%An example of an LLM-based agent is a travel agent tasked to “book a flight to a sunny place next week”. The perception module transforms the input for the LLM. The agent's brain module reasons for a good destination based on past user travels and weather reports, while the action module uses tools to book a flight and a hotel (cf. Sec. \ref{sec:relatedwork} for more details).

\emph{Challenges of LLM-based agents:} LLM-based agents, however, face quality issues. They are complex, distributed applications with diverse components. These systems must cope with (partial) failures of single or multiple components, network failures, data inconsistency, and more (e.g. \cite{cristian1991understanding} or \cite[Chap. 8]{kleppmann2017designing}). Additionally, the LLM, as a central component, is affected by systemic quality issues. A well-known issue is that LLMs generate outputs that are nonsensical or unfaithful to the provided source content (\textit{hallucination}~\cite{ji2023survey}) without telling the user. A real-world example involved a chatbot from a Canadian airline that promised discounts to customers, which were actually not available. The airline was later held liable (cf. \cite{BBC-air-canada}). Another quality issue is \textit{jailbreaking}, i.e., bypassing the LLM’s safety measures to provoke undesirable generation \cite{zou2023universal}. Consider, for example, an agentic system performing financial transactions. In such settings, erroneous actions can result in substantial financial losses. Furthermore, the manipulation of xAI’s Grok 4 in 2025 (e.g., \cite{TechTalks-grok}) highlighted the negative impact of jailbreaking.
% Additionally, \cite{deng2023masterkey} pointed out jailbreaking's negative impact in general 
Lastly, \cite{laban2025llmslostmultiturnconversation} showed that \textit{agent performance decreases in longer multi-turn conversations}. 

\emph{Optimization potentials of testing LLM-based agents:} since LLM-based agents are designed to support users with little or no supervision, they require extensive testing. Testing agents is currently focused on \textit{acceptance level} tests. In software engineering, acceptance testing determines if the completed software meets the customer's needs. Put simply, it probes whether the software does what the users want \cite{ammann2016introduction}. For LLM-based agents, the agent is viewed as a black box system and evaluated against both functional and non-functional requirements from the end-user's perspective. While tests at acceptance level are easy to define, code, and understand, they give \textit{limited insight into the internal behavior}, are \textit{time-consuming}, and thus, \textit{expensive to execute and evaluate}, \textit{difficult to automate}, and hence \textit{hard to scale} (similar to software acceptance testing \cite{fischbachautomaticcreationacceptancetests}). The black box view of the agent hinders the identification of internal root causes of unexpected behavior or failed acceptance tests, especially for more complex systems, such as conversational or multi-agent systems. Evaluating the agent's brain, the LLM, can involve a human, semantic comparison techniques such as metrics (e.g., cosine similarity) or LLM-as-a-judge \cite{zheng2023judging}. Since the result of these comparisons is a number ranging from -1 to 1, a threshold needs to be defined to evaluate an answer. Thus, building reliable LLM-as-a-judge systems is challenging (\cite{gu2024survey, zhao2025one}), evoking Juvenal's quote, “sed quis custodiet ipsos custodes?” [“but who will guard the guards themselves?”]. Additionally, the LLM's sensitivity to prompt variations \cite{jang2023can} impacts the agent's output. As a result, obtaining a specific agent behavior, e.g., invoking a particular tool, can be a try-and-repeat issue and hampers automatic checking of results. Lastly, testing conversational agents involves covering multi-turn conversations with numerous branches in various languages through comprehensive testing. 

\emph{Contribution:}
In this paper, we present methods to enable structural testing 
of LLM-based agents, bringing proven software engineering practices to this 
emerging domain. We make three key contributions:

\textbf{Enabling software engineering best practices for agents:} we demonstrate how structural testing enables the test automation pyramid, regression testing, test-driven development, and multi-language testing workflows for LLM-based agents.

\textbf{Technical methods for structural testing:} we use traces based on OpenTelemetry to capture agent trajectories, mocking to enforce reproducible LLM behavior, and assertions for automated verification. This enables testing of individual agent components and their interactions.

\textbf{Validation through case studies:} representative case studies demonstrate automated execution and faster root-cause analysis. These methods can reduce testing costs and improve quality by increasing coverage, enhancing reusability, and facilitating earlier defect detection.

We provide an open-source implementation under \href{https://www.apache.org/licenses/LICENSE-2.0}{Apache 2 license}, thereby facilitating unrestricted access and implementation. We hope this will spur adoption across the community. The source code is available via \href{https://github.com/awslabs/generative-ai-toolkit}{GitHub} and uses the \href{https://docs.aws.amazon.com/bedrock/latest/APIReference/API_runtime_Converse.html}{Amazon Bedrock Converse API}. Migration to other environments is fairly straightforward: classes for real and mocked LLM interface as well as the OpenTelemetry attributes have to be adapted to the new environment or LLM.
\section{Background and related Work} \label{sec:relatedwork}
\begin{figure}[htbp]
    \centering
    \includegraphics[width=0.48\textwidth]{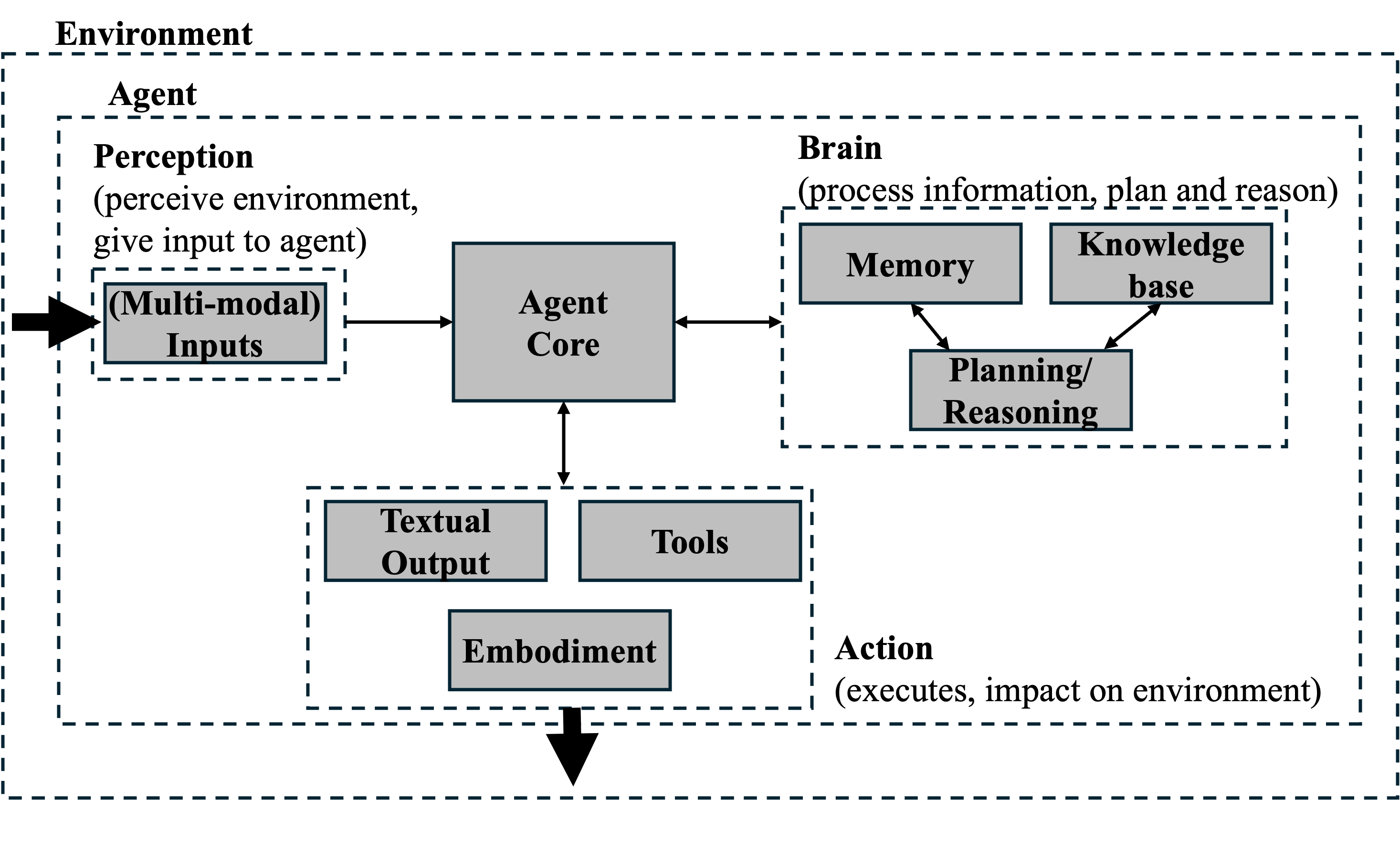}
    \caption{Structure of an LLM-based agent. Source: based on \cite{xi2023risepotentiallargelanguage}}
    \label{fig:UML-Agent-structure}
\end{figure}

\textbf{LLM-based agents:} can be deployed as single agents or multi-agent systems, where "multiple autonomous agents collaboratively engage in planning, discussions, and decision-making" \cite{guo2024largelanguagemodelbased}.
Overviews of historical and current LLM-based agents, their structure and components can be found in \cite{xi2023risepotentiallargelanguage,naveed2023comprehensive, guo2024largelanguagemodelbased, wang2024survey}. For the remainder of this paper, we will use the agent structure as shown in Fig. \ref{fig:UML-Agent-structure}.

%, i.e. several specialized agents interacting together.% \cite{guo2024largelanguagemodelbased} defines multi-agents as systems in which "multiple autonomous agents collaboratively engage in planning, discussions, and decision-making".%, thus offering "advanced compatibilities"\cite{weng2023llm}. 

%there are several publications with varying descriptions of the structures of LLM-based agents. %Some of the first publications were \cite{weng2023llm} and \cite{VarshneyAgent}.

Agents operate in a constant loop. The \textit{agent core} orchestrates this loop, coordinating between modules via interfaces or API and managing connections to tools, knowledge bases, (cloud-based) LLM, and other agents. The \textit{perception module} perceives changes in the external environment and then converts multi-modal information (often textual input) into an understandable representation for the agent. This allows agents to process textual, audio, and visual data. This information is forwarded to the \textit{brain module}. The brain module, which is primarily composed of an LLM, is the core of an agent. Its main functions are reasoning, planning, and processing available information. Reasoning and planning means breaking down a goal into steps or subtasks using a technique called ReAct \cite{yao2023reactsynergizingreasoningacting} or some variation thereof. For processing information, the brain can utilize memory (the agent’s past observations, thoughts, and actions with a user) and knowledge bases (external storage of typically very large amounts of information). Additionally, the memory is enriched with data generated by the brain for future use. The brain module generates sequences of actions which are forwarded to the \textit{action module}. The action module carries out the execution by providing textual output, using external tools (via functions registered with the agent or external functions via API) and embodiment (the capability to actively perceive, comprehend, and interact with physical environments and generate specific behaviors to modify the environment based on the LLM’s extensive internal knowledge). The constant loop allows the agent to continuously get feedback and interact with or impact its surrounding environment \cite[Chap. 3]{xi2023risepotentiallargelanguage}. 

\textbf{Quality issues of LLM:} include hallucinations, security vulnerabilities, sensitivity to prompts, performance degradation in multi-turn conversations, which can occur as a single or compound event. \cite{ji2023survey} surveys \textit{hallucinations} and defines it as "the generation of texts or responses that exhibit grammatical correctness, fluency, and authenticity, but deviate from the provided source inputs (faithfulness) or do not align with factual accuracy (factualness)". \cite{bender2021dangers} were among the first to publish about this characteristic, but called it stochastic parrot. \cite{ye2023cognitivemiragereviewhallucinations} give a taxonomy and \cite{tonmoy2024comprehensivesurveyhallucinationmitigation} show an overview of mitigation techniques. However, \cite{xu2024hallucinationinevitableinnatelimitation} and \cite{banerjee2024llmshallucinateneedlive} showed that hallucination is impossible to avoid completely. \textit{Jailbreaking} an LLM involves circumventing measures to prevent undesirable generation and was first shown by \cite{zou2023universal}. \cite{xu2024comprehensivestudyjailbreakattack} give an overview of possible jailbreaks. \cite{OWASPLLM2025} and \cite{MITRE-matrix} show other security risks for LLM-based applications.

\textbf{Quality issues of LLM-based agents:} The LLM's \textit{sensitivity to prompt variations} \cite{jang2023can} as well as longer multi-turn conversations \cite{laban2025llmslostmultiturnconversation} can impact the agent's performance. In multi-agent systems, the quality issues of LLM can quickly compound. For example, a common failure mode in agentic systems is planning and reasoning deficiencies whereby agents generate flawed execution plans due to, among other things, misinterpretation of user instructions, inadequate task decomposition, unspecific, ambiguous, and overlapping tool descriptions, or failure to anticipate dependencies between tools (\cite{cemri2025multi}, \cite{ning2024defining}). Another common way in which agentic systems break includes failure to determine task completion, leading to repetitive execution cycles (infinite loops) until forced termination \cite{cemri2025multi}. Finally, operational issues such as failing to call a tool with the correct parameter set or syntax can be a common cause of agent failures in production. In more complex multi-agent systems, another failure mode stems from inter-agent coordination issues, including task derailments or ignoring inputs of other agents. \cite[Chap. 6.3]{xi2023risepotentiallargelanguage} gives a literature survey over potential failures of agents with emphasis on security and trustworthiness, \cite{tian2023evil} focuses on safety. While these publications mostly cover single agents, \cite{cemri2025multi} defines a taxonomy for failures of multi-agents.

\textbf{Testing LLM-based agents:} given the shown quality issues, agents have to be tested extensively.  Research and development on testing agents currently focus on acceptance tests, which are rather an evaluation or validation than a verification ("are we building the right product?" versus "are we building the product right?", as stated by \cite{boehm1979software}). We discussed the benefits and challenges of accepting LLM-based agents in Section \ref{sec:intro}. \cite{chang2023surveyevaluationlargelanguage}, \cite{aleti2023softwaretestinggenerativeai}, and \cite{yehudai2025surveyevaluationllmbasedagents} give an overview for evaluating agents, \cite{ShamimSinghalQualityAssuranceTestingMultiAgent} focus on multi-agents and \cite{hua2024trustagentsafetrustworthyllmbased} on security.

\textbf{Frameworks for testing LLM-based agents:} LLM-based agents are an emergent technology, as are methods and frameworks used for their testing. In  \cite{kohl2024generativeaitoolkit} we listed frameworks for developing and operating agents such as \href{https://mlflow.org/}{MLflow}, \href{https://www.comet.com/site/products/opik/}{Opik}, or \href{https://github.com/awslabs/generative-ai-toolkit}{Generative AI toolkit}. For testing LLM-based agents, based on a non-exhaustive survey at the time of writing, the available tooling and frameworks can be categorized as follows: 
\begin{itemize}
    \item \textbf{Public cloud provider tooling:} \href{https://aws.amazon.com/bedrock/evaluations/}{Amazon Bedrock Evaluations}, \href{https://cloud.google.com/vertex-ai/generative-ai/docs/models/evaluation-overview}{Google Cloud GenAI evaluation service}, and \href{https://azure.microsoft.com/en-us/products/ai-foundry}{Microsoft Azure AI Foundry} are deeply integrated, tailored, and optimized into the provider’s offering. These products allow evaluating agents at the acceptance test level, but do not (yet) support structural testing of agents. % These products allow evaluating agents, but no testing on a deeper level than acceptance tests. 
    \item \textbf{Commercial off-the-shelf software:} there are several available offerings such as \href{https://github.com/confident-ai/deepeval}{DeepEval} by Confident AI, \href{https://langfuse.com/}{LangFuse}, \href{https://www.langchain.com/langsmith}{LangSmith}, \href{https://phoenix.arize.com/}{Phoenix Arise} and \href{https://wandb.ai/site/}{Weights and Biases}. These tools track the agent’s internal behavior and interactions, although they currently offer no testing on a deeper level than acceptance tests. \href{https://www.comet.com/site/products/opik/}{Opik} offers the concept of unit tests based on PyTest, though its focus is on the LLM pipeline. Additionally, these tools are not (completely) open source and not free of charge (but with a limited free product offering).
    
    \item \textbf{Free-to-use software:} \href{https://strandsagents.com/}{Amazon Strands Agents},
    \href{https://www.microsoft.com/en-us/research/project/autogen/}{Microsoft AutoGen},
    \href{https://mlflow.org/}{MLflow}, and \href{https://docs.ragas.io/en/stable/}{Ragas} are open source and free to use. These tools offer traces, but do not (yet) support testing on a level deeper than acceptance tests.
\end{itemize}

Additionally, even though several of these frameworks offer traces, there are currently only basic conventions for structure, attributes, or naming, but no standardization.

\textbf{Software testing:}  has been extensively researched with a substantial body of literature for several decades. \cite{myers1979} defined its purpose as "a successful test case ... detects an as-yet undiscovered error", but testing "can only show the presence of errors, not their absence"\cite{dijkstra1972humble}. Testing can be categorized into \textit{functional} (or black box-testing) and \textit{structural testing} (or glass box testing). "In functional testing, the program or system is treated as a black box. It is subjected to inputs, and its outputs are verified for conformance to specified behavior. The software’s user should be concerned only with functionality and features, and the program’s implementation details should not matter. Functional testing takes the user’s point of view"\cite[Chap. 2.2]{beizer1990software}. "Structural testing does look at the implementation details" as they are "aimed at exploring the consistency of a component’s implementation with its design" \cite[Glossary]{beizer1990software}. A detailed overview of testing techniques and approaches can be found in  \cite{beizer1990software} or \cite{ammann2016introduction}. Testing forms a significant part of the software development life cycle, with estimates ranging from  40\%\cite[Chap. 1]{sommerville2011software}, 50\% (\cite{myers1979}, \cite[pp. 126]{endres2003handbook}, \cite[Chap. 1.1]{beizer1990software}) or even 75\% \cite{hailpern2002software}. 

\textbf{Test automation:} given testing's importance, intensive research has been done to increase its efficiency, i.e., reduce test effort with limited or no impact on test quality (e.g., \cite{fewster1999software,graham2012experiences}). A popular approach to increasing test efficiency is the Test Automation Pyramid. Introduced by \cite[Chap. 16]{cohn2010succeeding}, most approaches use three layers but with different naming (e.g., \cite{MehtaGoogleTestPyramid2014} and \cite{WackerGoogleTestingBlog2015}). The lowest layer consists of tests for small software functions (\textit{unit tests}), which, if passed, are followed by testing the interactions of these small functions (\textit{integration tests}) and culminate in testing the entire system (\textit{acceptance tests}). While the concept of unit tests dates back to \cite{Benington1956}, \cite{cohn2010succeeding}'s idea was to unify different test levels and provide a guideline for distributing tests, favoring simple, automatable, low-level unit or integration tests. Since only a successful passing of all tests of one layer leads to testing the next layers, this concept helps to provide a good trade-off between test time and depth. Today, this pyramid is typically executed as part of an automated Continuous Integration/ Continuous Deployment (CI/CD, defined by \cite{fowler2006continuous}) pipeline. 

Several publications show case studies (e.g., \cite{kumar2016impacts},  \cite{garousi2017test}) or questionnaires for developer teams (e.g., \cite{kasurinen2010software}, \cite{wang2020software}) to identify benefits of test automation. \cite{rafi2012benefits} give a systematic literature review and point out that only a few publications show empirical evidence or exact figures for benefits, which include \textit{test reusability}, \textit{repeatability}, \textit{test coverage}, and \textit{effort saved in test executions}. \cite{patel2022state} give a survey of test automation in DevOps teams defining test automation as an essential enabler for DevOps with additional benefits of accelerating releases in higher quality and efficiency. Regarding quantitative results, \cite{dougherty2002test} reported a 40\% reduction in automation costs and a 50\% reduction in maintenance costs. \cite{amannejad2014search} use return-on-investment calculation to show significant impact when test artifacts are reused several times (up to 307\% for test design, 675\% for test execution, and 41\% for test evaluations if tests are used more than 10 times). \cite{kumar2016impacts} show effects as factors for COCOMO (Constructive Cost Model, \cite{boehm1979software}).

\textbf{Test-driven development:} as introduced by \cite{beck2022test}, is an iterative procedure to produce "clean code that works". It starts with writing a small, automated test for a functionality that fails, then changing the code to pass the test, and then concluding by refactoring the code.
Hence, tests are written before developing a feature- in contrast to other traditional testing strategies where tests are usually built after completing the implementation. Some publications see benefits of TDD as to improve code coverage and quality, documentation, regression testing and developer productivity (e.g., \cite[Chap. 8.2]{sommerville2011software}, \cite{buchan2011causal}, \cite{scanniello2016students}, \cite{shull2010we}), while others see mixed results such as impact on quality but little or no positive \cite{Rafique2013TDD} or even detrimental effects (\cite{canfora2006evaluating, dogvsa2011effectiveness}) on productivity. \cite{ghafari2020research} point out that these contradicting results are partly due to how case studies were defined, e.g., unclear scope, lack of developer experience with TDD, focus on building new software instead of maintaining, as well as not analyzing long-term benefits. \cite{mock2024generative} show how to use LLM to speed up the TDD workflow, while \cite{Mathews2024} detail how to improve the quality and correctness of software code generation by an LLM via TDD. \cite{ChavezLTTDLLM} and \cite{Krawczyk-tdd-agents2024} showed how to adapt test-driven development for an LLM in general, while \cite{PatelGuardrails2024} adapted TDD for the specific use case of guard railing in- and outputs.  These approaches all evaluate an agent on the acceptance level.
%\section{Structural testing of LLM-based agents} \label{sec:contribution}
\section{Technical building blocks} \label{sec:contribution}

In this section, we detail the core contribution of this paper. Starting with techniques to enable structural testing, we then describe these tests for agent components (i.e., \textit{unit tests}) and their interactions (i.e., \textit{integration tests}). The techniques are adapted from software engineering. We provide a brief description of each technique and its benefits, followed by a detailed explanation of its application to the agent domain.
%the domain of LLM-based agents.

\subsection{Methods for structural testing}

\subsubsection{Traces as key enabler} 
Tracing enables detailed observability of a system, facilitates root cause analysis and bug fixing, and helps to understand the implemented system. Traces record and store information and events happening during the execution of a software or system. They are used extensively in complex systems, such as embedded or distributed systems \cite{moc2001understanding}, and are generated by frameworks like \href{https://opentelemetry.io/}{OpenTelemetry} or Dapper~\cite{sigelman2010dapper} (see \cite{janes2023open} for an overview). In the OpenTelemetry tracing model, a unique \textit{trace\_id} is assigned for each incoming request. Every single action to fulfill this request (e.g., calling a function) is recorded in the form of a data structure called \textit{span} consisting of attributes such as span\_id, name, start, and end timestamp, as well as detailed information about the action.

For LLM-based systems, several frameworks for developing (cf. \cite[Sec. 2]{kohl2024generativeaitoolkit}) or testing agents (cf. Sec. \ref{sec:relatedwork}) offer traces, mostly based on OpenTelemetry. These frameworks allow capturing the agent's full trajectory to handle a user input. This allows monitoring deployed agents and observing the impact of the LLM's varying output on the whole agent. Additionally, tracing agents facilitate incident analyses, for instance, in cases of customer complaints or unexpected agent outputs, such as invoking more tools per request than before or replying with "I cannot help you with that" when not expected. %Typically, in distributed systems, an incoming (HTTP) request is "a trace" and all actions happening while serving this request are recorded as a so-called "span" under that same trace, e.g. a database query or calling an external API. 

For agents, we also utilize the OpenTelemetry span model, where one unique trace covers one user input and one agent response, i.e., one turn. Each turn has spans for significant operations such as accessing memory, invoking LLM, or using tools. We also use a fixed attribute, ai.conversation.id, to group traces, facilitating to find and analyze all relevant traces for a single conversation. Traces are stored in a database or files.

\subsubsection{Automating tests} tracing not only helps to identify the root cause of a failure, but can also be used to test the system or agent itself by comparing the expected behavior of a system with the real behavior as captured by traces. In software engineering, assertions are used to compare expected with real behavior. In the event of discrepancies, the program terminates with an error. 
Sample frameworks for this include \href{https://jestjs.io/}{Jest} and \href{https://docs.pytest.org/en/}{PyTest}, which allow for defining human-readable test cases that include detailed information about the error. Assertions enable developers to detect bugs early, as code execution stops upon a failed assertion. 

We use them to compare the expected behavior of agent components and their interactions against real behavior as captured by traces. Hence, this allows us to test agents on a deeper level than acceptance tests. A longer conversation between a human and an agent consists of multiple traces (one per turn), including multiple spans of actions the agent took during that turn.  Hence, finding the relevant pieces can be a cumbersome task. This can be facilitated using regular expressions, string search functions, or classes that filter for the relevant trace or span data for the assertion. Additionally, building such classes provides the opportunity to write test cases that are easy to understand. 

Taken together, this enables running and checking tests via PyTest in an automated workflow. Additionally, these tests can be integrated into a CI/CD pipeline, enabling automated test runs for each change. This also prevents deploying  faulty agents, as the pipeline stops in case of a failed assertion. For agent components with deterministic behavior, defining and checking structural tests works the same as in software engineering. However, the LLM's varying output makes testing the LLM and its interactions with other agent components more challenging: testing a specific behavior of an agent, such as calling a tool, can be a time-intensive trial-and-error process.

\subsubsection{Mocking to facilitate tests involving an LLM} In software engineering, mocks are used to facilitate testing of complex or non-deterministic objects, especially in specific situations or environments, such as reproducible testing of connection errors of distributed systems. Mocks are "objects pre-programmed with expectations which form a specification of the calls they are expected to receive" \cite{FowlerMocks2007} and are based on the proxy pattern \cite{gamma1995design}. They offer the same external interfaces as the objects they replace, but their pre-definable responses ensure deterministic behavior. This allows for faster tests, e.g., simulating a database instead of setting up a real one.  Exemplary mocking frameworks are \href{https://github.com/google/googletest/}{Google Test} for C++, \href{https://jmockit.github.io/}{JMockit} for Java, \href{https://docs.python.org/3/library/unittest.mock.html}{Mock} for Python, or \href{https://www.postman.com/}{Postman} for API. 

Mocking an LLM is very similar to mocking a component in software engineering. Agent frameworks using a cloud-based LLM invoke the LLM via API (e.g. \href{https://docs.aws.amazon.com/bedrock/latest/APIReference/API_runtime_Converse.html}{Amazon Bedrock Converse API}) and receive as a response an action (textual output or tool to call, cf. Fig. \ref{fig:UML-Agent-structure}). Mocking an LLM, then, means building an object with the same interface and possible return values, but without actually calling an LLM. The agent then calls the mock and receives user-definable responses. Hence, mocking allows enforcing specific LLM behavior and thus facilitates testing by providing reproducible test conditions for the LLM itself as well as its interactions with the other agent components. Furthermore, mocking offers fast testing as  mocked LLM return instantly. Finally, using a mocked LLM saves costs compared to cloud-based LLM, which are priced on a per-token basis.

\subsection{Structural testing of agent components} \label{subsec:agent-components}

Table \ref{tab:unit-tests} shows examples of issues with agent components we want to cover with tests.

\begin{table}[ht]
\small
\centering
\begin{tabular}{lll}
\textbf{Component} & \textbf{Test rationale} & \textbf{Sample test} \\
\cline{1-1} \cline{2-2} \cline{3-3}

Agent           & initialization works?     & Agent w/ valid model    \\ 
core            & valid configuration?      & id successfully created?  \\ \hline

                & exists? accessible?       & write, then read data \\ \cline{2-3}
Memory          & ensure privacy            & access to memory of    \\
                &                           & other users possible? \\ \hline

                & empty/ faulty             & Test empty/ long strings \\
Knowledge       & queries                   & incl. special characters  \\ \cline{2-3}

base            & data updated              & add data, then new data   \\
                & or augmented             & returned on querying  \\ \hline

LLM             & does fine-tuned           & invoking fine-tuned LLM   \\ 
                &  LLM work?                & w/ sample prompt returns \\ 
                &                           & plausible results  \\ \hline

Tools           & return plausible          & get\_weather(Munich)  \\ 
                & output(s)?                & returns plausible values \\ \hline

Textual         & validate user             & Test injections, long input \\ 
input           &   input                   & and incompatible formats \\ 

%input           & System prompt         & sys. prompt integrated  \\
%                & integration           & and formatted correctly \\ 

\end{tabular}
\caption{Examples for possible tests of agent components}
\label{tab:unit-tests}
\end{table}

A software unit is "the smallest testable piece of software...[consisting] of several hundred or fewer lines of source code", i.e., a component, module, or a single behavior of software.  \cite[Chap. 3.7]{beizer1990software} defines unit testing as "the testing we do to show that the unit does not satisfy its functional specification and/ or that its implemented structure does not match the intended design structure". Unit tests work by entering different parameters and checking the expected behavior of the component itself in isolation \cite[Chap. 8.1.1]{sommerville2011software}. 

This works for all agent components with deterministic behavior like tools, memory, or knowledge bases. The LLM, however, has varying output and offers only limited insight. Therefore, we need traces to gain insight into its actions and to test expected  against actual behavior, as captured by traces.

\subsection{Structural testing of interactions of agent components} \label{subsec:agent-interactions}

%Fig. \ref{fig:UML-Agent-sequence} depicts possible agent interactions and table \ref{tab:integration-tests} shows exemplary issues of these interactions we want to cover with tests.

Table \ref{tab:integration-tests} illustrates possible issues of interactions between agent components we want to cover with tests.

\begin{table}[ht]
\small
\centering

\begin{tabular}{lll}
\textbf{Component} & \textbf{Test rationale} & \textbf{Sample test} \\
\cline{1-1} \cline{2-2} \cline{3-3}
Planning and    & history persistent    & Conversation continues? \\
memory          & after restart?         & Follow-up queries work?\\ \hline

Planning and    &  check if both     & querying company policy  \\
knowledge       &  interact correctly?       & retrieves correct data\\ \hline

Agent core      & extracted right      & "Get weather in ..." calls   \\
and tools       & tool and parameter?   & right tool and parameters\\ 
\end{tabular}
\caption{Examples for possible tests of agent interactions}
\label{tab:integration-tests}
\end{table}

Tests for interactions between components are called integration tests, which are "specifically aimed at exposing the problems that arise from the combination of components" \cite[Chap. 3.7]{beizer1990software}. Their purpose is to “explore how components interact with each other and with data under the assumptions that the components and objects they manipulate have all passed their local tests.” \cite[Glossary]{beizer1990software}. Testing interactions of agent components with deterministic behavior is equivalent to integration testing in software engineering, and traces are not necessary. But traces are especially handy when we want to test the interaction of the LLM with other components or the interactions of multiple, distributed agent components (e.g., a model in the cloud). 

Mocking is useful for making integration tests faster or simpler (e.g., using a mock instead of a complex database). It’s possible to mock all interacting components, but then the integration test does not really provide insights into the integration. Hence, the extent of mocking should be chosen carefully. However, testing of interactions between LLM and other components is different, due to the LLM’s varying output. Here, mocking enforces reproducible behavior of the LLM.
\section{Best practices and workflows} \label{sec:application}

In this section, we show how structural testing can be applied to facilitate regular workflows for testing agents. 

\subsection{Facilitate test workflows for agents}

\subsubsection{Regression tests} 
LLM-based agents should be tested after any change affecting their components (e.g., a changed prompt, updates to the knowledge base, new or changed tooling, an LLM update, etc.) to ensure continued correct functionality. Structural tests allow testing changed components themselves and their interactions with other parts of the system specifically. An example would be testing if a newly added company policy in the agent's knowledge base is returned correctly. Developers can then test the knowledge base itself and then the interaction of the agent's planning module with the knowledge base.

\subsubsection{Root cause analyses}
Structural tests can be used to drill down on specific agent components or interactions, facilitating root cause analyses in cases of failed acceptance tests, unexpected behavior, or bugs. An example would be if the agent failed an acceptance test to book a flight and a hotel. Here, structural tests can help to pinpoint the issue, e.g., wrong credentials, flight not available, or overbooked. 

\subsubsection{Testing different languages} structural testing -especially integration tests- ease testing the agent's performance in different languages, since we can check via traces if the agent did the right action instead of checking for a semantically correct answer in the correct language, as required at the acceptance level. An example would be testing a user query searching for events in a city. Testers can run the equivalent queries "show me events in Munich", "\begin{CJK*}{UTF8}{gbsn}搜慕尼黑的活\end{CJK*}" (Chinese) or "Zeige mir Veranstaltungen in München" (German) and just have to check the traces for an invocation of the correct tool and city as parameter. Hence, these tests can be run and evaluated automatically. 

\subsubsection{Testing multi-turn conversations} multi-turn conversations are conversations between an agent and a human or other agents involving multiple exchanges. Given the LLM's and thus the agent's varying output, this means covering multiple possible branches with tests. With mocking to "enforce" LLM-behavior and functions to switch between mocked and real agent, developers can steer a conversation to test possible branches of multi-turn conversations. Traces then allow detailed insight into the agent’s actions. Additionally, if a test's pass or fail result can be checked by traces, it can be run and evaluated automatically in a CI/CD pipeline.

\subsection{Test automation pyramid for LLM-based agents} \label{subsec:test-automation-pyramid}

Functional and structural testing "target different kinds of bugs. Functional tests can, in principle, detect all bugs, but would take infinite time to do so. Structural tests are inherently finite but cannot detect all errors, even if completely executed" \cite[Chap. 2.2]{beizer1990software}. Exhaustive functional tests could theoretically detect all possible bugs, but exponentially growing input combinations limit real world feasibility to small subset of inputs. Structural tests are finite, as the code they test has a limited number of statements and branches. However, these tests cannot detect unimplemented or incorrectly implemented features. Hence, effective testing should combine both, low-level structural and user-level functional testing.

\begin{figure} [ht]
    \includegraphics[width=\columnwidth]{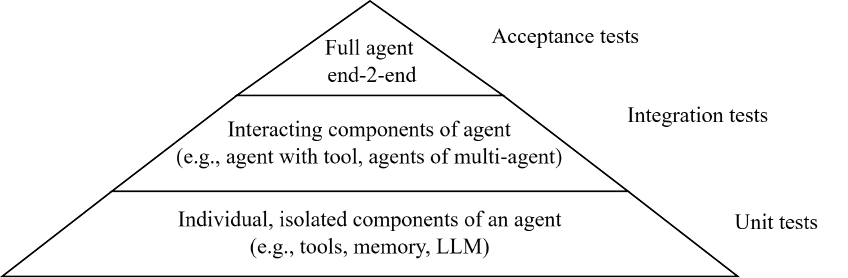}
    \caption{Test framework for LLM-based agents.}
    \label{fig:test-automation-LLM}
\end{figure}

The test automation pyramid (cf. Section \ref{sec:relatedwork}) is a well-known approach in software engineering for combining functional and structural tests, with a preference for structural testing. Fig. \ref{fig:test-automation-LLM} shows an adaptation to the domain of agents (with the corresponding layers of the original pyramid on the right side).

Let’s consider an easy example for the testing pyramid for agents. Suppose a user is interacting with an agent and wants to learn about things to do in the city of Munich in the coming week. The user initiates the following query: "show me events in Munich next week". To answer this query, the agent is expected to call a dedicated tool for searching events, called \textit{search\_events()} with parameters \textit{Munich} and a specific \textit{date}. We start the pyramid with testing the \textit{agent's components in isolation} (\textit{unit tests} in the test automation pyramid). Sample tests check if the tool \textit{search\_events()} is callable and the results are as expected. After tests show the agent's components work in isolation as intended, we test the \textit{interactions of a component with other components} (\textit{integration tests} in the test automation pyramid). We regard single agents of a multi-agent system as components whose interactions are tested with integration tests. %On this level, we are also more interested in return type and structure than in exact return values for sake of automation. 
An example for a test would be the interactions of the agent with the \textit{search\_events(city, date)} tool. As the LLM's output can vary, we mock the LLM to enforce reproducible behavior. This ensures, for instance, that an integration test can confirm that the agent invoked a specific tool with correct parameters. %Mocking also facilitates testing multi-turn user - agent conversations. Here, dynamically switching between mocked and real agent allows us to reach and test specific branches of a conversation. 
If these tests pass as well, we test the \textit{full agent end-to-end} from an end-user perspective (\textit{acceptance tests} in the test automation pyramid). End-to-end means we want to test the whole environment an agent is embedded in (e.g., the agent could be part of a software solution, or controlled by speech). In our example, this could be the initial query sent as a JSON message via HTTPS to a chatbot deployed as a container in the cloud.

Similar to software engineering, the test pyramid for agents can be run as part of a CI/CD pipeline. This allows efficient test runs even with many tests per layer, as only a successful passing of all tests of one layer leads to testing the next layers (“fail-fast” strategy). We run the pyramid for each new agent release, which is triggered by changes such as updated prompts, models, or data. This software engineering best practice prevents the deployment of faulty agent releases into the production environment.% We also run the pyramid regularly to ensure a stable system.

\subsection{Test-driven development for LLM-based agents}

Structural testing is a key enabler for adapting TDD (cf. Sec. \ref{sec:relatedwork}) to the domain of LLM-based agents. %A very simple example for this would be testing if an agent can call an unregistered tool. In TDD's first step, \textit{write a failing test} developers build the agent and the tool, but do not register the tool with the agent. Then they write a test case with a user prompt having the agent invoke the unregistered tool. As the agent does not know the tool, it cannot invoke the tool. With acceptance testing this can be trial-and-repeat. With mocking the agent's behavior can be enforced and traces allow to simplify the evaluation of the test by checking the agent's traces for an invoke of the tool. For TDD's next step, \textit{make test pass}, developers simply have to register the specific tool with the agent, run the test again and check the traces again. In TDD's third step, \textit{refactoring}, \TDO{add refactoring steps}
An example would be a customer complaint caused by an agent executing an incorrect action or responding with "I'm sorry, I cannot fulfill that request." In TDD's first step, \textit{write a failing test}, developers write a test case that reproduces the complaint using input from its traces. If the complaint occurred as part of a multi-turn conversation or other specific conditions, mocking the agent helps to facilitate reproducing these circumstances. In TDD's next step, \textit{make the test pass}, developers can verify that their solution works by running tests against the agent's traces. In TDD's last step, \textit{refactoring}, developers improve the code without changing its external behavior against tests. This can involve removing duplicates, adding comments, changing variable names, or splitting up code into separate methods.
\section{Case studies} \label{sec:case_studies}

We applied methods and workflows to several case studies. Due to limited space, we show only two representative examples. The code shown in this section uses Amazon Bedrock Converse API and the \href{https://github.com/awslabs/generative-ai-toolkit}{Generative AI Toolkit}. We provide full source code of another use case, a simple travel agent, detailing methods, framework, and CI/CD pipeline via this \href{https://github.com/awslabs/generative-ai-toolkit/tree/main/examples/sample_agent}{link}. 

\subsection{Driver assistance agent}
We developed a proof-of-concept for an LLM-based agent  assisting drivers with questions about their vehicle or in need of support. With this PoC, we wanted to find out if an LLM-based agent can cover or even speed up some complex, time-intensive interactions between a driving customer and a hotline in multiple languages in high quality, e.g., retrieving and updating customer account data, (pre-) diagnosis of the vehicle on user request, or customer questions about vehicle features. Although this use case focuses on vehicles, it also covers standard topics related to helplines for other technical products.

%Architekturbild/ technische Implementierung
\begin{figure}[ht]
    \centering
    \includegraphics[width=0.46\textwidth]{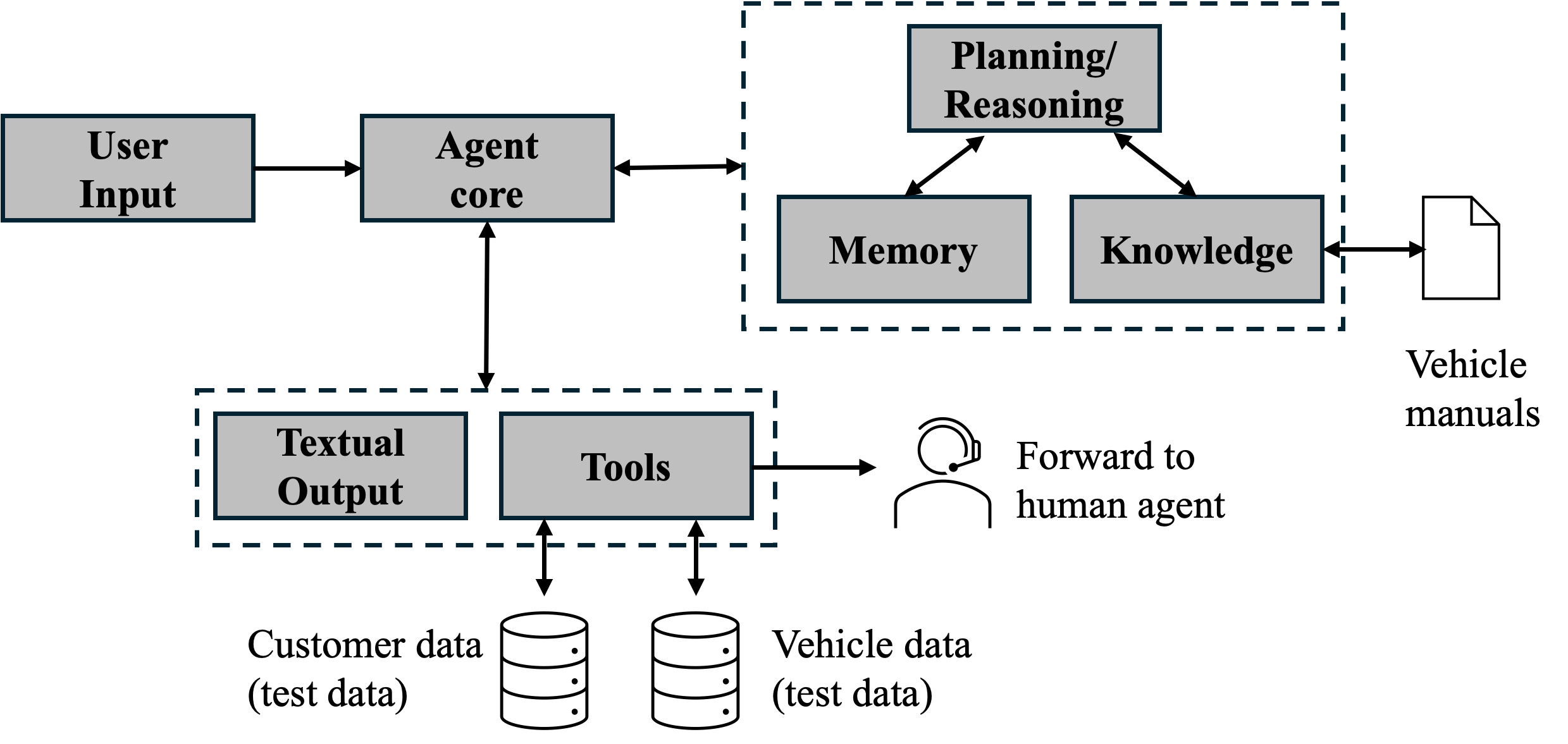}
    \caption{Simplified architectural overview for driver assistance agent.}
    \label{fig:Architecture-RSA-agent} 
\end{figure}

Figure \ref{fig:Architecture-RSA-agent} illustrates the agent's simplified structure. For this PoC, we use a RAG (Retrieval-augmented generation, \cite{lewis2020retrieval}) to store vehicle documents, such as manuals and tools, interacting with external databases for customer and vehicle data.

Listing \ref{lst:RSA_update_data} shows an often-occurring example of customers wanting to change their data. We define and run the test case, using the \textit{Expect} class to verify in the traces that the correct tool was called. %For this test, we are mainly interested in if the agent understood the request, identified the right tool to change the data, extracted the correct parameters, and calls the tool to update the data. 
We can test this in different languages by translating the multi-line string below the first comment.
\begin{lstlisting}[caption={Test changing customer data},captionpos=b, label=lst:RSA_update_data, language=python]
def test_change_customer_data(rsa_agent):  
    test_case = Case(
        user_inputs=[
            "<Start conversation>"\
            "<PhoneNr>+555-12345</PhoneNr>",
            # to test other languages, 
            # translate multi-line string below
            "Hi, I am John Doe." \
            "My new phone number is +555-98765."\
            "Could you please update my data?"],)
    traces = test_case.run(rsa_agent)
    Expect(traces).tool_invocations\
        .to_include("get_customer_information")
    Expect(traces).tool_invocations\
        .to_include("update_customer_information")\
            .with_input({
                "ucid": "1", 
                "phoneNr": "+555-98765",})
\end{lstlisting}
In listing \ref{lst:RSA_book_appointment}, we want the agent to book an appointment. This time, we use a mocked LLM for the first part of the conversation so that we can steer the conversation towards a desired state for our test. Switching between a mocked LLM and a real LLM (via \textit{add\_real\_response}) allows for scripting a specific situation and then testing the agent’s reaction.
\begin{lstlisting}[caption={Test conversation to book appointment for winter tires},captionpos=b, label=lst:RSA_book_appointment, language=python]
def test_change_tire_type(mocked_LLM, rsa_agent):
    # agent greets user, asks for permission to scan
    mocked_LLM.add_output(
        text_output=[
            "I am a prototype for"\
            "a driver Assistance agent."\
            "May I diagnose your vehicle?"]) 
    #  get customer and vehicle info, scan vehicle
    mocked_LLM.add_output(
        tool_use_output=[{
                "input": {"vin": "XXX"},
                "name": "get_vehicle_status",
            },{
                "input": {"phoneNr": "+555-98765"},
                "name": "get_customer_information",
            },])        
    # agent's response to results of scan
    mocked_LLM.add_output(
        text_output=[
            "Everything is fine."\
            "Can I help you further?"])
    # we invoke real LLM for each text_output above
    mocked_LLM.add_real_response()
    mocked_LLM.add_real_response()
    # run test case with user_input, test case
    # will use prepared, mocked responses above
    traces = Case(
        user_inputs=[
            "Hello, can you help me?", 
            "Yes, go ahead with the diagnosis",
            "I need winter tires."\
            "Would next Monday work?",],)\
                .run(rsa_agent) # run test case
    # no need to test the agent's text response,
    # since we test if right tools were invoked.
    # assert tool get_vehicle_status was invoked
    Expect(traces).tool_invocations\
        .to_include("get_vehicle_status")\
            .with_output({
                "found": True,
                "status": {
                    "lastUpdate": "2025-08-28",
                    # vehicle data cut for brevity
                    "....": "....",},})
    # assert tool invocations with right parameters
    Expect(traces).tool_invocations\
        .to_include("list_available_appointments")
    Expect(traces).tool_invocations\
        .to_include("book_appointment")\
            .with_input({
                "appointment_id": "IX94",
                "reason": "install winter tires"})
\end{lstlisting}

\begin{figure}[ht!]
    \centering
    \includegraphics[width=0.464\textwidth]{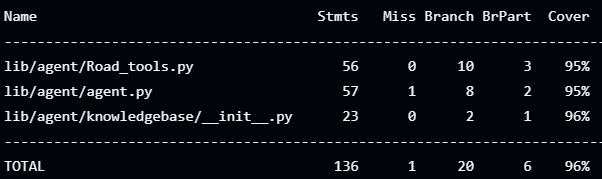} 
    \caption{screenshot of driver assistance agent's test coverage}
    \label{fig:CICD-code_coverage} 
\end{figure}

We also utilize test coverage in the CI/CD pipeline. Test coverage is a class of metrics measuring how much of the code is executed during test runs (cf. \cite[Chap. 2.3.4 and 2.3.5]{beizer1990software}, \cite{ISO29119}). For the pipeline in Fig. \ref{fig:CICD-code_coverage}, we chose \textit{statement} (each statement executed at least once) and \textit{branch coverage} (each branch executed at least once). These metrics enable our developers to identify which parts of an agent have been thoroughly tested and where to add additional tests.

Running a CI/CD pipeline allows for automated test runs and thus reduces test efforts and costs. The test pyramid's "fail-fast" strategy offers to further reduce these costs, as we only test on higher levels if all tests of a previous layer passed. 

%As the test results can be checked via traces, this allows automated test runs and thus reduces test efforts and costs. The "fail-fast" strategy of the test pyramid offers to reduce test time and costs further, as we only test on higher levels if all tests of a previous layer passed. 

\subsection{Cloud incident root cause analysis agent}

\begin{figure}[ht]
    \centering
    \includegraphics[width=0.48\textwidth]{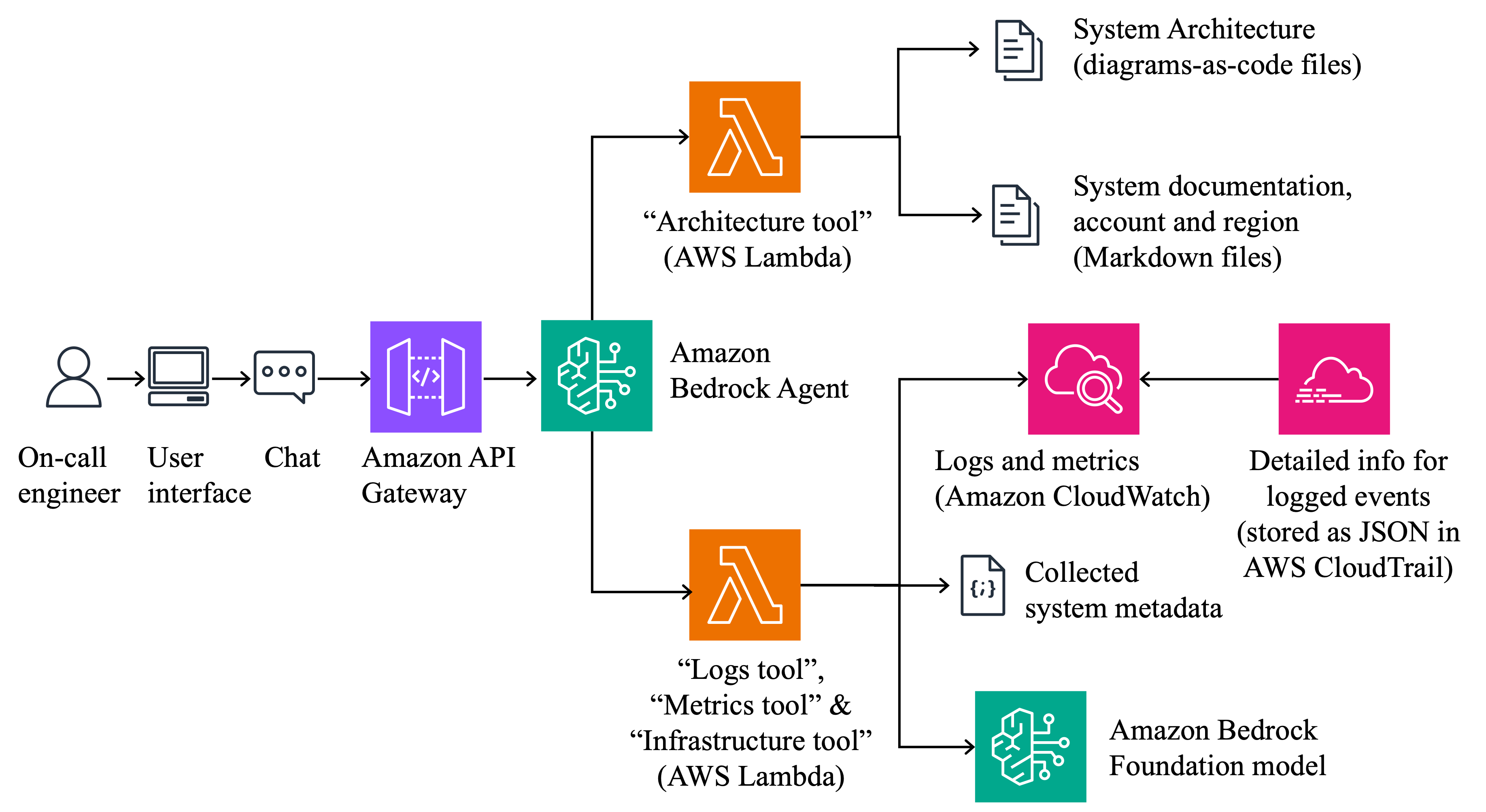}
    \caption{Architectural overview for root cause analysis agent. Source: \cite{RCA-agent}}
    \label{fig:Architecture-RCA-agent} 
\end{figure}

The second case study involves an LLM-based agent supporting root cause analyses of service incidents in the cloud. The agent assists engineers with system understanding and real-time insights, thus significantly reducing resolution times \cite{RCA-agent}. Given an incident, an engineer enters a description of the issue into the agent via a chat interface. The agent's capabilities are modeled on tasks a real-world engineer would do to solve issues. The agent builds hypotheses for the root cause based on the technical documentation of the cloud account (such as architectural diagrams and Markdown files), system logs and metrics, and its knowledge. The engineer can then instruct the agent to give more details for the defined hypotheses or to build new ones. Additionally, the engineer can ask the agent to create sample code to fix the issue. Fig. \ref{fig:Architecture-RCA-agent} shows an architectural diagram of the agent. 

Structural testing helps us facilitate the test workflows. Due to limited space, we provide some sample test cases instead of including the full source code.
The first example uses traces and structural tests to verify that the agent is utilizing its tools in the correct sequence. Hence, we can ensure that the agent first runs the architecture tool to understand the architecture and services used, before running other tools, such as log and metric tools, to gather more detailed information. This is important as errors for authorization and authentication (e.g., accessDenied), resource-related (e.g., ResourceNotFoundException), or throttling are similar for multiple AWS services. Thus, directly analyzing error logs without referencing the actual architecture can lead to invalid hypotheses, additional iterations, increased time for root cause analysis, and lower user acceptance. The second example demonstrates using mocks to simplify the test environment. Instead of setting up test accounts with specific failure patterns, we can save time by mocking the return values of the agent tools. We can then check the traces to see if the return values of mocked tools lead the agent to identify the scripted root cause, e.g., removing or changing a security group ingress rule, which can cause connectivity issues between services (a simple yet often occurring incident in real life). The third example is a fault pattern with failure propagation to other services. We want to test that the agent identifies the system architecture and then walks back the entire failure propagation, as described in the logs, to detect the failure pattern’s beginning and its root cause. The last example is an update of documents (e.g., architecture diagrams or docs), indicating a change of the systems to be diagnosed. Thus, we can test if the agent is still functional and to avoid stale test cases. 

We can then save time and money by running regression tests for the agent’s interactions with the updated systems, instead of repeating full test runs. While these examples take some time to build, they are useful in the long term, as they facilitate tests for refactoring and future changes to the system, reducing maintenance costs.

\subsection{General findings}

For a thorough quantitative analysis of the benefits of structural testing it is necessary to conduct a large-scale study, including comparison with or without such tests (similar to effects of testing in software engineering, cf. Sec. \ref{sec:relatedwork}). 

However, upon applying the methods to several use cases, we observed several qualitative benefits for developers. With traces, our developers have \textit{higher observability} of the agents, as traces give detailed insight into the agent’s full trajectory and show more details than the agent's outputs (e.g., how often were tools called? Were tools called in sequence or parallel? How many cycles did the agent run?). We utilize traces in both development and production, enabling us to identify divergences quickly and easily. Additionally, traces offer a \textit{fast root cause analysis} of failed tests (including acceptance tests). Using assertions on traces in a CI/CD pipeline helps to prevent pushing faulty agents into production, as failed assertions stop the pipeline run. This enables \textit{higher quality} from the first deployments. A multi-turn conversation can have dozens of traces, each consisting of multiple spans with lots of attributes. Here, the Expect classes enable easy analysis of traces and spans, facilitating the construction of tests that are easier to understand and therefore more maintainable and reusable. Mocking helps our developers build reproducible tests involving LLM and their interactions. Together, this makes for tests that are fast and easy to develop, as well as quick to run. Hence, they help our developers test frequently during development and aim towards \textit{high test coverage} with minimal impact on development costs. And since tests of agent components can be reused even across different use cases, they also \textit{reduce costs for building and maintaining tests}. Given the pace of innovation in the domain, deployed LLM-based agents are likely to undergo changes. Structural tests enable rapid and straightforward testing of modified components and their impact on the system, thereby reducing test time. Additionally, our developers can run the test automation pyramid regularly to test for a stable system. The pyramid structure represents an efficient tradeoff, emphasizing many fast, cheap unit and integration tests over fewer expensive acceptance tests. 

% Finally, structural testing offers test-driven development for agents; for some of our developers a preferred way to develop. 

%Using the same tracing in production and development facilitates reproducing complex or sporadic failures from the field and testing fixes - even in different languages.
\section{Conclusion and future work}
\label{sec:conclusion}

Testing LLM-based agents presents unique challenges due to their varying behavior, complex multi-turn interactions, and black-box nature. In this paper, we presented methods to enable structural testing of LLM-based agents, addressing these challenges through three key techniques: traces to capture agent execution, mocking to enforce reproducible LLM behavior, and assertions for automated verification.

We demonstrated how these methods facilitate the testing of agents and enable the adaptation of best practices from software engineering to the domain of agents, such as the test automation pyramid. Our case studies demonstrate that structural testing facilitates automated test execution, faster root-cause analysis, and higher test coverage. Qualitative observations from applying these methods across multiple use cases suggest potential for reduced testing costs through increased test reusability and earlier defect detection.

We acknowledge two limitations. First, our evaluation is qualitative; a large-scale quantitative study comparing development with and without structural testing would provide stronger evidence of benefits. Second, our implementation uses Amazon Bedrock Converse API as a unified interface to multiple LLM. However, a migration to other environments or LLM is confined to adapting the mocked and real interface. 

%Second, while our implementation uses Amazon Bedrock Converse API, the underlying methods (based on OpenTelemetry and standard mocking patterns) are platform-agnostic. Hence, a transfer to other LLM environments is straightforward.

Given the observed benefits, we have integrated these methods into our standard agent development framework and open-sourced the implementation under Apache 2 license at \href{https://github.com/awslabs/generative-ai-toolkit}{GitHub}. We encourage the community to adopt, extend, and evaluate these methods in diverse contexts.

Future work includes extending the methods to additional LLM platforms, conducting quantitative studies on testing effectiveness and cost savings, and developing more comprehensive sample applications that demonstrate the full testing pyramid.

%Future work includes extending the methods to additional LLM platforms, more quantitative studies on testing effectiveness and cost savings. Finally, we are investigating if structural tests on agent trajectories, captured by traces, can accelerate the self-evolving of agents and enhance the quality of the reinforcement learning reward signal for general tasks (instead of LLM-as-a-judge).

% Future work includes extending the methods to additional LLM platforms, conducting quantitative studies on testing effectiveness and cost savings. Finally, we are investigating if structural tests on agent trajectories, captured by traces, can accelerate the self-evolving of agents and enhance the quality of the reinforcement learning reward signal for general tasks.
\section*{Acknowledgements}
The authors would like to thank (in alphabetical order) Adam Raymer, Dustin Hughes, Fabrizio Avantaggiato, Felix Willnecker, Florian Seidel, Jonas Schroeder, Luisa-Sophie Gloger, Mariano Kamp, Mihai Radulescu Kober, Paul Weber, Raghvender Arni, Raphael Perri, and Rui Costa for their support, insightful inputs, comments, feedback, and reviews. The authors used OpenAI’s and Anthropic’s language generation models to generate parts of the text, which were subsequently reviewed, edited, and revised.

\bibliographystyle{IEEEtran}
\balance
\bibliography{main}

\end{document}